\documentclass[aps,preprintnumbers,twocolumn,prb,amsmath,amssymb,superscriptaddress]{revtex4}

\usepackage{graphicx}
\usepackage{dcolumn}
\usepackage{bm}
\usepackage{color}
\usepackage{hyperref}

\graphicspath{{Figures/}}

\begin{document}
\title{Skyrmion morphology in ultrathin magnetic films}
\author{I.~Gross}
\affiliation{Laboratoire Charles Coulomb, Universit\'{e} de Montpellier and CNRS, 34095 Montpellier, France}
\affiliation{Laboratoire Aim\'{e} Cotton, CNRS, Universit\'{e} Paris-Sud, ENS Cachan, Universit\'{e} Paris-Saclay, 91405 Orsay Cedex, France}
\author{W.~Akhtar}
\affiliation{Laboratoire Charles Coulomb, Universit\'{e} de Montpellier and CNRS, 34095 Montpellier, France}
\author{A. Hrabec}
\affiliation{Laboratoire Charles Coulomb, Universit\'{e} de Montpellier and CNRS, 34095 Montpellier, France}
\affiliation{Laboratoire de Physique des Solides, CNRS UMR 8502, Universit\'{e}s Paris-Sud et Paris-Saclay, 91405 Orsay Cedex, France}
\author{J.~Sampaio}
\affiliation{Laboratoire de Physique des Solides, CNRS UMR 8502, Universit\'{e}s Paris-Sud et Paris-Saclay, 91405 Orsay Cedex, France}
\author{L.~J.~Mart\'{\i}nez}
\affiliation{Laboratoire Charles Coulomb, Universit\'{e} de Montpellier and CNRS, 34095 Montpellier, France}
\author{S.~Chouaieb}
\affiliation{Laboratoire Charles Coulomb, Universit\'{e} de Montpellier and CNRS, 34095 Montpellier, France}
\author{B.~J.~Shields}
\affiliation{Department of Physics, University of Basel, Klingelbergstrasse 82, Basel CH-4056, Switzerland}
\author{P.~Maletinsky}
\affiliation{Department of Physics, University of Basel, Klingelbergstrasse 82, Basel CH-4056, Switzerland}
\author{A.~Thiaville}
\affiliation{Laboratoire de Physique des Solides, CNRS UMR 8502, Universit\'{e}s Paris-Sud et Paris-Saclay, 91405 Orsay Cedex, France}
\author{S.~Rohart}
\affiliation{Laboratoire de Physique des Solides, CNRS UMR 8502, Universit\'{e}s Paris-Sud et Paris-Saclay, 91405 Orsay Cedex, France}
\author{V.~Jacques}
\affiliation{Laboratoire Charles Coulomb, Universit\'{e} de Montpellier and CNRS, 34095 Montpellier, France}
\email{vincent.jacques@umontpellier.fr}

\begin{abstract}

Nitrogen-vacancy magnetic microscopy is employed in quenching mode as a non-invasive, high resolution tool to investigate the morphology of isolated skyrmions in ultrathin magnetic films. The skyrmion size and shape are found to be strongly affected by local pinning effects and magnetic field history. Micromagnetic simulations including a static disorder, based on the physical model of grain-to-grain thickness variations, reproduce all experimental observations and reveal the key role of disorder and magnetic history in the stabilization of skyrmions in ultrathin magnetic films. This work opens the way to an in-depth understanding of skyrmion dynamics in real, disordered media.
\end{abstract}
\date{\today}

\maketitle

Current-induced motion of magnetic textures in ultrathin films is the cornerstone of innovative applications in spintronics such as the racetrack memory~\cite{parkin2008magnetic}.  However, in technologically relevant magnetic materials, structural defects often result in considerable pinning that limits the propagation velocity. Whereas domain walls necessarily experience all the disorder landscape when propagating along a magnetic track~\cite{lemerleprl1998,metaxas}, skyrmions, localized magnetic quasiparticles, are predicted to move in two dimensions while avoiding strong pinning sites~\cite{Fert2013}. Skyrmions are thus expected to display limited interaction with disorder, leading to highly efficient motion at low current densities. Surprisingly, several recent experiments have instead shown that skyrmion dynamics are in fact strongly affected by disorder~\cite{jiang2016direct,litzius2016skyrmion,hrabec2016current,legrand2017room,Boulle_arxiv,zeissler2017}, suggesting that pinning effects have been oversimplified in seminal simulations of skyrmion dynamics~\cite{iwasaki2013disorder,sampaio2013nucleation}. These observations motivate a more precise description of disorder in magnetic materials hosting skyrmions.
\begin{figure}[t]
\includegraphics[width = 8.2cm]{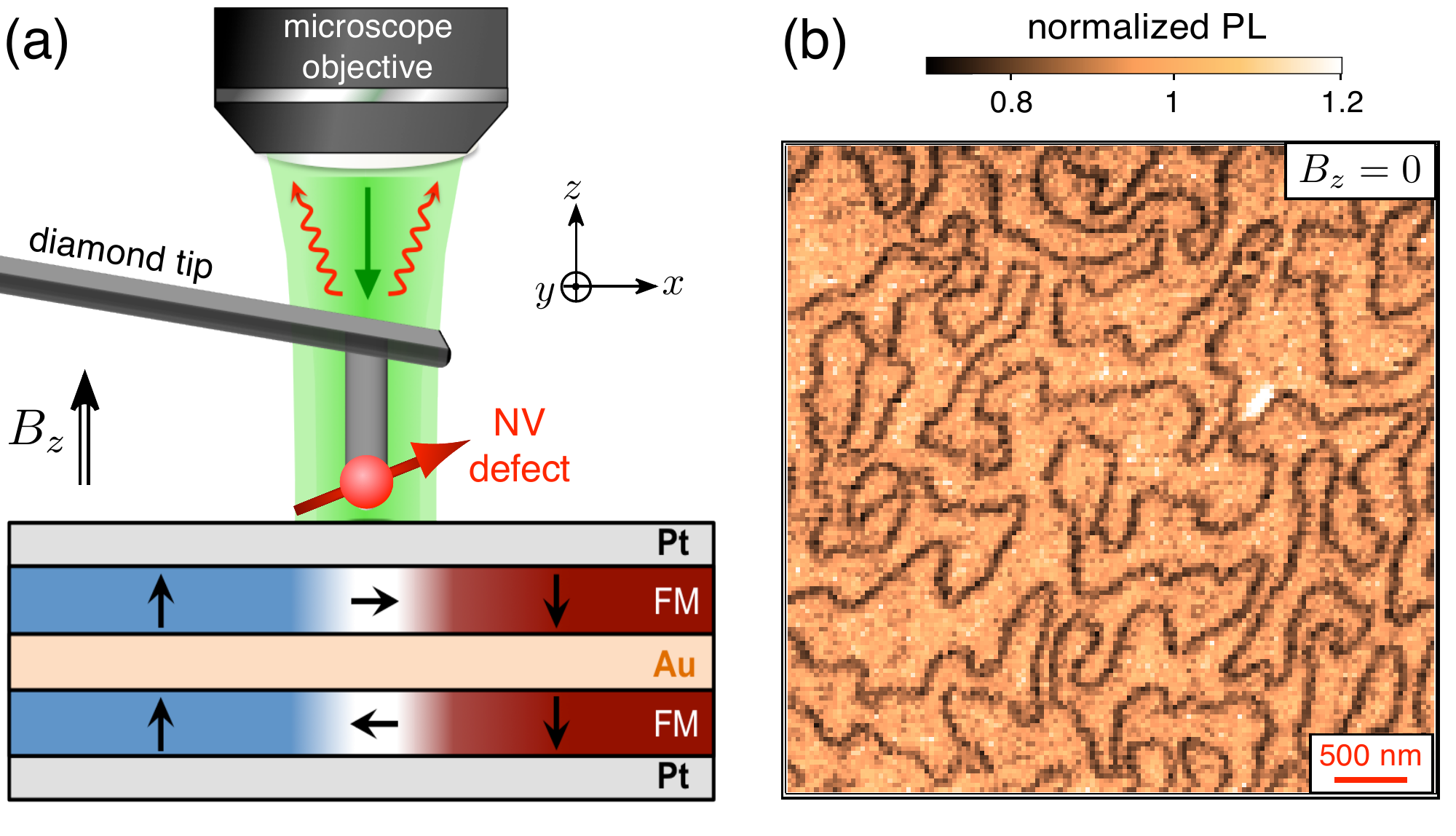}
\caption{(a) Principle of the experiment. A single NV defect placed at the apex of a diamond tip is employed as a non-invasive, scanning nanomagnetometer operating under ambient conditions.
A microscope objective is used both to excite (green arrow) and collect the magnetic-field dependent photoluminescence (PL) of the NV defect (red wavy arrows). The magnetic sample is a symmetric bilayer system with a stack of Pt(5nm)$\backslash$FM$\backslash$Au(3nm$\backslash$FM$\backslash$Pt(5nm), where FM$=$Ni(4\AA)$\backslash$Co(7\AA)$\backslash$Ni(4\AA). (b) PL quenching image recorded in zero external field ($B_z=0$).}
\label{Fig1}
\end{figure}

  \begin{figure*}[t]
\includegraphics[width=18cm]{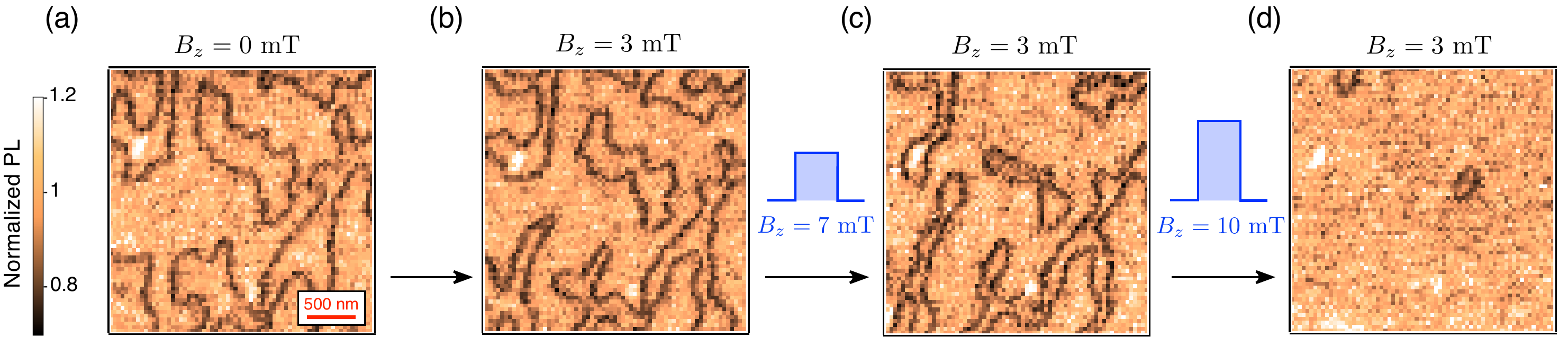}
\caption{Step-by-step formation of isolated skyrmions by applying an external out-of-plane magnetic field $B_z$. (a) PL quenching images recorded at zero field and (b) $B_z=3$~mT. (c,d) Images recorded at $B_z=3$~mT after applying a $10$-s field pulse of (c)~$7$~mT and (d)~$10$~mT.
The bright PL spots correspond to particles on the sample which serve as position references.}
\label{Fig2}
\end{figure*}

Whereas average magnetic parameters of ultrathin films - magnetization, anisotropy, Dzyaloshinskii-Moriya interaction, damping - can be easily measured, the effects of structural disorder remain highly challenging to evaluate and model. In this work, we investigate the impact of disorder on the size and shape of isolated skyrmions. Such a study requires high quality imaging of the skyrmion morphology.
Room temperature skyrmion imaging has been achieved by numerous experimental techniques, mostly using transmission microscopes (transmission electron microscopy\cite{pollard2017observation,McVitie2017} and scanning transmission X-ray microscopy\cite{moreau-luchaire2016,woo2016observation,zeissler2017,Boulle_arxiv}), photoemission electron microscopy \cite{boulle2016} and scanning probe techniques (mostly magnetic force microscopy (MFM) \cite{hrabec2016current,legrand2017room,yagil2017,bacani2016} and nitrogen-vacancy (NV)  center magnetometry\cite{Skyrmion_Harvard}).
However, transmission techniques lack in sensitivity and require the use of samples with several magnetic layers\cite{moreau-luchaire2016,woo2016observation}, while magnetic force microscopy, which can be used for thinner samples, induces perturbation and therefore may modify the skyrmion morphology. In contrast, NV  center magnetometry has recently emerged as a high sensitivity, perturbation-free technique to probe spin textures in ultrathin films \cite{Rondin_2012,Rondin_2013,Rondin2014,tetienne2014nanoscale,tetienne2015nature,Skyrmion_Harvard}.
Here, we use scanning NV magnetometry in \textit{quenching mode}~\cite{Rondin_2012} to obtain high spatial resolution images of magnetic skyrmions in a sample relevant for spintronic applications. Our measurements are carried out in ambient conditions and without perturbing the magnetic structure of the skyrmions. The recorded distributions of skyrmion size and shape reveal the key role of disorder and magnetic history in stabilizing isolated skyrmions. Micromagnetic simulations including disorder, modeled as a grain to grain thickness fluctuation, allow an accurate description of all experimental observations, opening the way to an in-depth understanding of the skyrmion dynamics in ultrathin films.

The sample used in this study is a symmetric magnetic bilayer system, which has shown skyrmion stabilization at room temperature under moderate external magnetic field, with state-of-the art skyrmion motion under current~\cite{hrabec2016current}. It consists of two ferromagnetic layers (FM$=$Ni$\backslash$Co$\backslash$Ni with a total thickness of 1.5~nm), separated by a 3~nm thick gold spacer, and sandwiched between two 5~nm thick Pt layers [see Fig.~1(a)]. The Dzyaloshinskii-Moriya interaction (DMI) arising at the Pt$\backslash$FM interfaces, combined with flux-closing dipolar fields~\cite{bellec2010}, promotes the stabilization of superimposed skyrmions having identical topological charge and opposite chirality in each FM layer. The resulting skyrmion pairs are strongly coupled by dipolar fields and behave as magnetic quasiparticles, which are hereafter referred to as skyrmions, for simplicity. Using MFM, it has been shown that such magnetic textures can be moved efficiently by electrical current with a critical current of about $2.5\times10^{11}$~A/m$^2$, reaching velocities as high as $60$~m.s$^{-1}$. The observation of a transverse deflection induced by the Magnus force~\cite{hrabec2016current}, an effect often referred to as the {\it skyrmion Hall effect}~\cite{jiang2016direct,litzius2016skyrmion}, has further confirmed the topological charge of chiral magnetic skyrmions.

Here we investigate the morphology of isolated skyrmions by using a single NV defect in diamond as an atom-sized magnetic field sensor~\cite{Maze2008,Balasubramanian2008,Rondin2014}. This technique, which can provide non-invasive and quantitative magnetic field measurements with nanoscale spatial resolution, has recently emerged as a versatile tool that offers valuable information on technologically relevant magnetic materials~\cite{tetienne2014nanoscale,tetienne2015nature,thiel2016quantitative,gross2016direct,Seagate2016,Skyrmion_Harvard,Du195}.
Combining optical illumination and microwave excitation, static magnetic fields are usually measured by recording the Zeeman-shift of the NV defect electronic spin sublevels though optical detection of the magnetic resonance (ODMR)~\cite{Rondin2014}. Such a measurement protocol becomes highly challenging for magnetic fields larger than $10$~mT with a significant field component perpendicular to the NV spin quantization axis. In this moderate-field regime, any off-axis magnetic field induces spin state mixing, leading to a drastic reduction in ODMR contrast~\cite{tetienne2012magnetic}. This situation is inevitably reached as soon as the NV sensor is brought in close proximity to a ferromagnet, {\it i.e.} when high spatial resolution is required. As an example, magnetic simulations indicate a stray field amplitude larger than $20$~mT at a distance $h=50$~nm above a domain wall in the bilayer magnetic sample studied in this work. For such fields, the ODMR contrast vanishes and quantitative magnetic field imaging cannot be performed with NV-based magnetometry. However, it has been shown that the decreased ODMR contrast is accompanied by an overall reduction of the NV defect photoluminescence (PL) intensity~\cite{Lai_2009, Epstein_2005}. This magnetic-field-dependent PL quenching can be exploited to map high magnetic field regions without the need of microwave excitation~\cite{tetienne2012magnetic}. Although not fully quantitative, we show below that this all-optical imaging mode is ideally suited to study the morphology of ferromagnetic textures with high spatial resolution.

As sketched in Fig.~1(a), we employ a single NV defect located at the apex of a nanopillar in a diamond scanning-probe unit~\cite{Maletinsky_2012, Appel_2016}. Once integrated into a tuning-fork-based atomic force microscope (AFM), this device enables scanning of the NV sensor in close proximity to the sample. For the present study, a calibration process following the method described in Ref.~\onlinecite{Hingant_2015} indicates a probe-to-sample distance of $\sim50$~nm. Magnetic field imaging is performed in the quenching mode by recording the NV defect PL intensity while scanning the magnetic bilayer sample. A typical PL quenching image recorded at zero field is shown in Fig.~\ref{Fig1}b. Sharp dark areas with a contrast greater than $\sim20$\% reveal regions of high stray magnetic fields, which correspond to domain walls organized in a worm-like structure. Although very similar images could be obtained with MFM~\cite{hrabec2016current}, the key advantage of NV-based magnetometry is the absence of magnetic back-action on the sample, which provides unambiguous field measurements.

\begin{figure*}[t]
\includegraphics[width = 18cm]{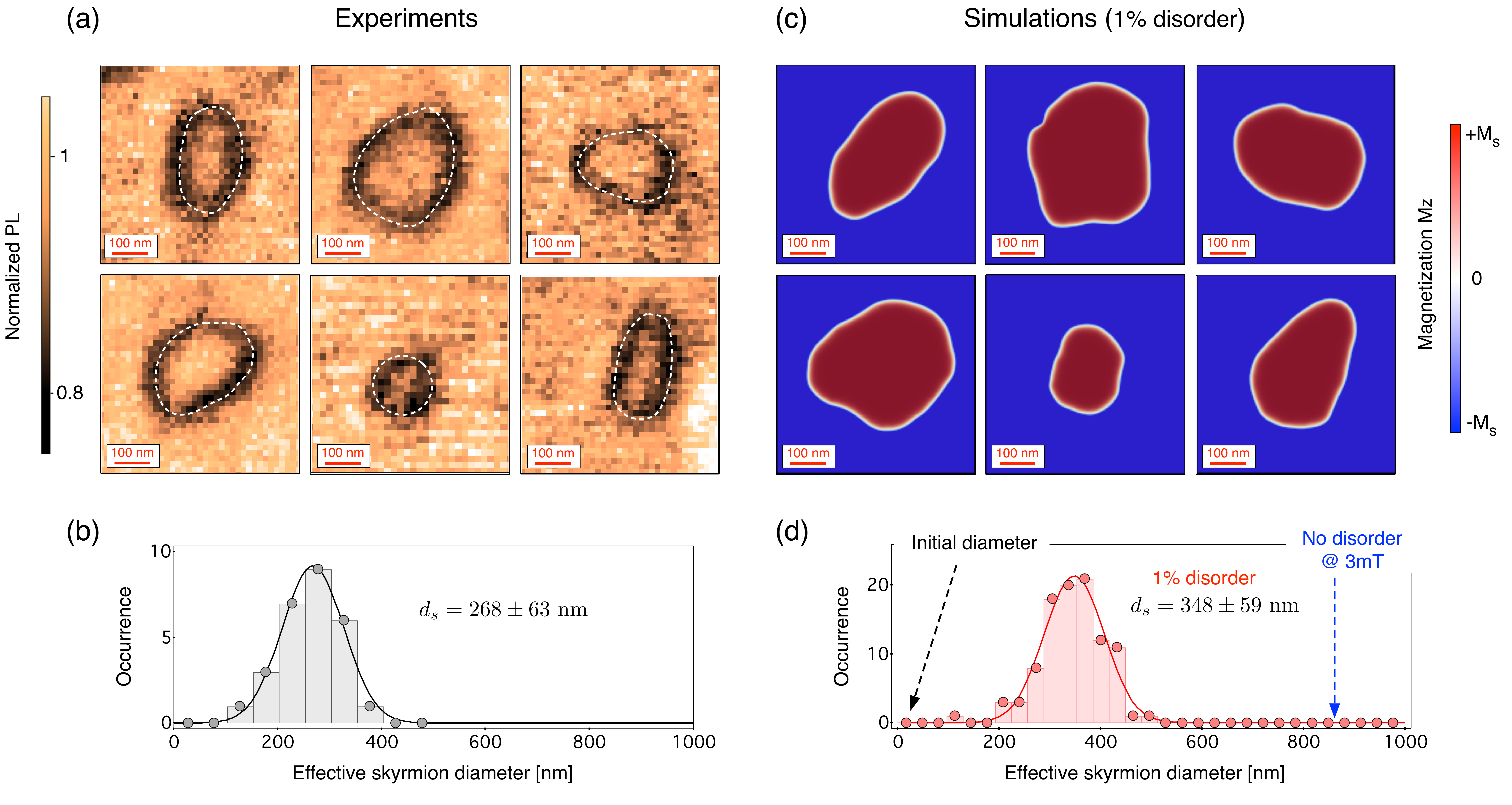}
\caption{(a) PL quenching images recorded above several isolated skyrmions at various position in the sample.
These experiments are performed at $B_z=3$~mT after applying a field pulse of $10$~mT.
The white dashed contours indicate the PL quenching ring from which the skyrmion area $\mathcal{A}$ is extracted. (b) Histogram of the effective skyrmion diameter $d_s$ extracted from measurements over a set of $27$ skyrmions. The solid line is a fit with a Gaussian distribution. (c) Typical micromagnetic simulations of the skyrmion spin texture by including thickness fluctuations with a relative amplitude of $1\%$. (d) Histogram of the effective skyrmion diameter obtained for a large number of randomly picked disorder configurations. The blue dashed arrow indicates the skyrmion diameter for a disorder-free sample ($860$~nm) and the black dashed arrow shows the initial skyrmion size in the simulation ($40$~nm) before expansion in a 3 mT field.
}
\label{Fig3}
\end{figure*}

Starting from a worm-like magnetization structure, isolated skyrmions are obtained by applying an external magnetic field $B_z$ perpendicular to the sample [Fig.~2]. For $B_z=3$~mT, one domain type starts to shrink but skyrmions are not yet formed [Fig.~2(b)]. In the next step, a $\sim$10~s magnetic field pulse of larger amplitude is applied in order to release domain walls from pinning sites. The magnetic field image is subsequently recorded at $B_z=3$ mT. As shown in Fig.~2(c), a field pulse of $7$~mT considerably compresses the magnetic domains, while after a $10$~mT pulse the wormy domain structure completely collapses, leading to the formation of isolated skyrmions [Fig.~2(d)]. In this experiment, the characteristic skyrmion size remains much larger than the NV-to-sample distance ($\sim 50$~nm). As a result, domain walls from opposite sides of the skyrmion can be easily resolved, leading to a dark ring in the PL quenching image. As mentioned above, the skyrmionic nature of such magnetic bubbles was proved in a previous work by the detection of the skyrmion Hall effect~\cite{hrabec2016current}.

Magnetic skyrmions were extensively imaged over the sample in order to obtain their size and shape distributions. Whereas perfectly circular skyrmions would be expected owing to the in-plane symmetry of the magnetic energy, we observe significant distortions of the magnetic texture [see Fig.~3(a)]. As analyzed below, these observations can be well explained by pinning effects induced by disorder in the sample. The characteristic skyrmion size is inferred by measuring the area $\mathcal{A}$ enclosed by the dark ring observed in the PL images. The effective diameter $d_s$ is then defined as $d_s=2\sqrt{\mathcal{A}/\pi}$, corresponding to a conversion of the distorted skyrmion geometry into a perfectly rounded shape. Measurement of $27$ isolated skyrmions leads to $d_s=268\pm63$~nm [Fig.~3(b)].

In order to understand these results, micromagnetic simulations including disorder were carried out with the MuMax3 code~\cite{Vansteenkiste2014}. The sample is modeled as two magnetic layers, each $t_0=1.5$~nm thick, with a 3~nm thick spacer. We use magnetic parameters extracted from previous measurements~\cite{hrabec2016current}: interfacial anisotropy $K_\mathrm{s}=0.75$~mJ.m$^{-2}$, saturation magnetization $M_{\mathrm{s}}=0.85$~MA.m$^{-1}$, exchange constant $A=12$~pJ.m$^{-1}$ and interfacial DMI constant $D_s=\pm0.3$~pJ.m$^{-1}$ with a negative (resp. positive) sign in the bottom (resp. top) FM layer. Disorder is included by a random fluctuation of the FM layer thickness $t$ between columnar grains arranged in a Voronoi fashion~\cite{Reichhardt2015,jue2016,pham2016,kim2017current}. The average lateral grain size is fixed to $15$~nm, as observed by high-resolution atomic force microscopy (AFM) imaging~\cite{SMnote}, and the thickness variation between grains is assumed to follow a normal distribution. Since the micromagnetic code requires a computational cell with a constant thickness $t_0$ over the whole sample, the saturation magnetization is varied from grain to grain as $M_\mathrm{s}t/t_0$. Averaged over the thickness, the uniaxial anisotropy $K$ and the effective DMI constant $D_{\rm eff}$ are also directly modified in each grain, {\it i.e.} $K=K_{\rm s}/t$ and $D_{\rm eff}=D_{\rm s}/t$.

In a disorder-free medium, magnetic simulations lead to circular skyrmions with a diameter $d_0=860$~nm under a 3~mT field, which is much larger than the experimental observation. However, measurements are performed after a 10-mT field pulse, which means that skyrmions are imaged after a compression of their diameter, followed by relaxation at 3~mT. To account for such a magnetic history, round skyrmions with a diameter $d_i=40$-nm, corresponding to the skyrmion size for a 10-mT field in a disorder-free medium, are first generated and then relaxed in a 3-mT field while including disorder. Figure~3(c) shows typical results of the simulation for thickness fluctuations with a relative amplitude of $1\%$. The calculated skyrmion morphology is similar to that observed in the experiments, revealing the key role of disorder and magnetic history in the stabilization of isolated skyrmions. Statistics over a large number of disorder distributions leads to an effective skyrmion diameter $d_s^{[1\%]}=348\pm59$~nm [Fig.~3(d)]. A similar analysis performed for disorder amplitudes of $2\%$ and $0.5\%$ thickness fluctuations leads to $d_s^{[2\%]}=136\pm55$~nm and $d_s^{[0.5\%]}=577\pm44$~nm, respectively [Fig.~4(a)]. Note that for grain size larger than the domain wall width $\Delta=\sqrt{A/K_\mathrm{eff}}$ (with $K_\mathrm{eff}=K-\frac{1}{2}\mu_0M_{\rm s}^2$ the effective anisotropy), the relaxation is unaffected by the grain size.\cite{SMnote}

It is striking to note that the impact of disorder on the skyrmion morphology is very strong, even for the relatively small thickness fluctuations considered. The key parameter fixing the skyrmion size is the domain wall energy $\sigma$, which involves all magnetic parameters~\cite{rohart2013}. For a single magnetic layer, $\sigma = 4\sqrt{AK_\mathrm{eff}}-\pi D_{\rm eff}$. Note that in the sample studied in this work, the bilayer behavior makes the relation slightly different~\cite{SMnote}, with no simple analytical formulas~\cite{hrabec2016current}. Media adapted for skyrmion stabilization are generally optimized to display a low effective anisotropy in order to decrease the domain wall energy to few mJ.m$^{-2}$. As a result, tiny fluctuations of any micromagnetic parameter are converted into large relative fluctuations of the effective anisotropy and domain wall energy. As an example, $1\%$ thickness fluctuations lead to about $30\%$ fluctuations in effective anisotropy and domain wall energy, respectively~\cite{SMnote}. Such large fluctuations explain the observed strong impact on the skyrmion spin texture.

Although illustrative, these micromagnetic simulations were performed at zero temperature, {\it i.e.} without including the effects of thermal fluctuations. As a result, a direct comparison between simulations and experiments does not allow extracting exact information about thickness fluctuations in the sample. These simulations rather provide a disorder model suitable for further 0~K simulations as used in previous studies~\cite{legrand2017room,kim2017current}. Since thermal agitation helps to overcome energy barriers induced by pinning sites, the actual thickness fluctuations in the sample are most likely larger than $1\%$. Micromagnetic simulations can be performed at finite temperature using an additional fluctuating field~\cite{Vansteenkiste2014,brown1963} but cannot accurately include all its impact~\cite{SMnote,grinstein2003} if the temperature is too large. Therefore, we have also investigated the relaxation process at 100~K, and found an agreement with the experiments for a thickness fluctuation of about 5~\%, a value closer to the AFM estimated roughness~\cite{SMnote}.

\begin{figure}[t]
\includegraphics[width = 7.8cm]{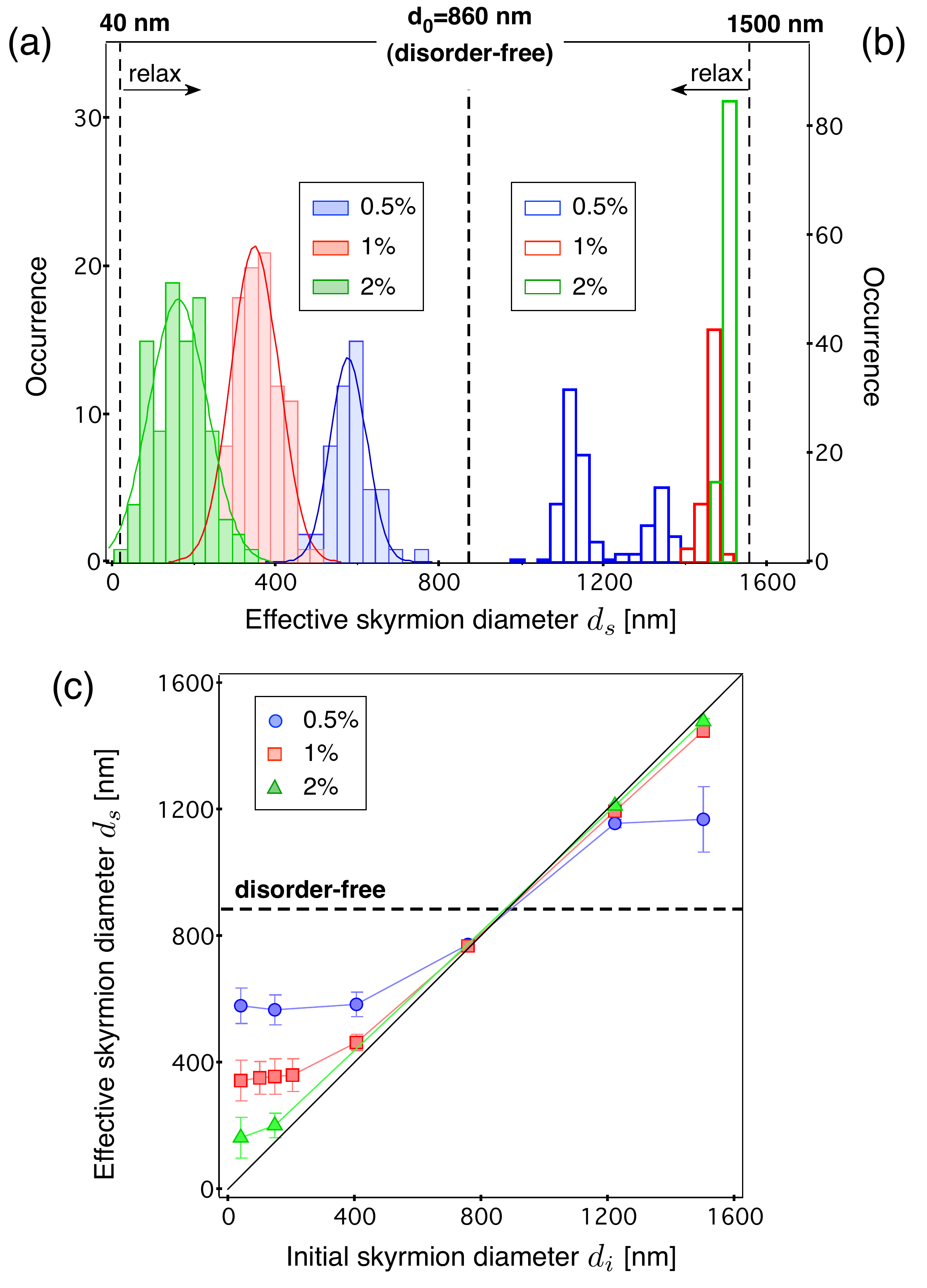}
    \caption{(a,b) Effective skyrmion size distribution obtained for different strengths of disorder after relaxation in a 3 mT field while starting with a skyrmion diameter of (a) 40 nm and (b) 1500~nm.
The black dashed line in the middle indicates the skyrmion diameter for a disorder-free sample ($860$~nm). (c) Effective skyrmion size after relaxation vs. initial size for different disorder strengths.
The error bars represent the variance of the size distribution. The black dashed line corresponds to a disorder-free sample and the diagonal (black solid line) is a guide to the eye for the case without relaxation.}
    \label{Fig4}
\end{figure}

To further illustrate how the skyrmion size depends on disorder and magnetic field history, the skyrmion relaxation was simulated under the same 3 mT magnetic field while starting from a large skyrmion with a diameter $d_i=1500$-nm. The resulting skyrmion size distribution is shown in Fig.~4(b) for various amplitudes of the thickness fluctuations. We now observe that the skyrmion size is larger than the one expected in a disorder-free sample. Such a measurement procedure was not accessible experimentally, as round 1500-nm diameter skyrmions could not be stabilized before the application of the 3~mT magnetic field. However, a qualitative comparison can be made by comparing the first two images in Fig.~2, where the $1400\times500$~nm$^2$ domain at the center in zero field relaxes under 3~mT into a $1200\times400$~nm$^2$ domain.

The disorder-limited relaxation is highlighted by plotting the variation of the effective skyrmion size obtained after relaxation ($d_s$) as a function of the initial size ($d_i$) [Fig.~4(c)]. In a disorder-free medium, the final size does not depend on the initial size and no magnetic history effect is found [see black dashed line in Fig.~4(c)]. Including disorder in the model always leads to $d_s<d_0$ (resp. $d_s>d_0$) when $d_i<d_0$ (resp. $d_i>d_0$).
Moreover, the simulations show that the skyrmion relaxation becomes more and more efficient when the disorder decreases, as expected. We finally note that for the smallest initial size, relaxation is always found more efficient than for larger initial size. This illustrates that the skyrmion energy variation versus its diameter is strongly asymmetric~\cite{hrabec2016current,boulle2016,rohart2013,buttner2016}, so that the force restoring the equilibrium size is stronger when $d_s < d_0$ compared to $d_s > d_0$.

In conclusion, we have used scanning-NV magnetometry in quenching mode to investigate the impact of disorder and magnetic history on the morphology of skyrmions in an ultrathin magnetic sample relevant for spintronic applications. A simple model of disorder based on thickness
fluctuations has been shown to provide a good description of the obtained results. This work opens the way to a detailed understanding of the dynamics
of skyrmions in real, disordered media \cite{jiang2016direct,litzius2016skyrmion}. Indeed, the uniformity of the skyrmion size directly influences skyrmion dynamics \cite{legrand2017room}, as the dissipation term in the Thiele equation grows with skyrmion size, whereas the gyrotropic term is independent, fixed by topology \cite{hrabec2016current}. Besides providing new insights into the impact of structural disorder on the morphology of magnetic skyrmions, this work also highlights the unique potential of NV magnetometry in quenching mode to study ferromagnetic textures with high spatial resolution under ambient conditions.

\noindent {\it Acknowledgements:} We thank J.~P. Tetienne and T.~Hingant for experimental assistance at the early stages of the project and K. Bouzehouane for experimental assistance with AFM measurements. This research has been supported by the European Research Council (ERC-StG-2014, {\sc Imagine}) and by the French Agence Nationale de la Recherche (ANR-14-CE26-0012, {\sc Ultrasky}).

\begin{widetext}

\vspace{0.2cm}

\begin{center}
{\large SUPPLEMENTARY INFORMATION}
\end{center}

\section{Grain size measurement}

The grain size in our sample has been estimated from the topographic atomic force microscopy image shown in Figure~\ref{fig_afm}. The average grain size is 13.6 nm with a full width at half maximum (FWHM) of 7 nm. The average roughness is 1.54 nm, which corresponds to about 7 \% of the total thickness (magnetic layer and non magnetic layer thickness).

\begin{figure}[ht]
\includegraphics[width = 0.45\textwidth]{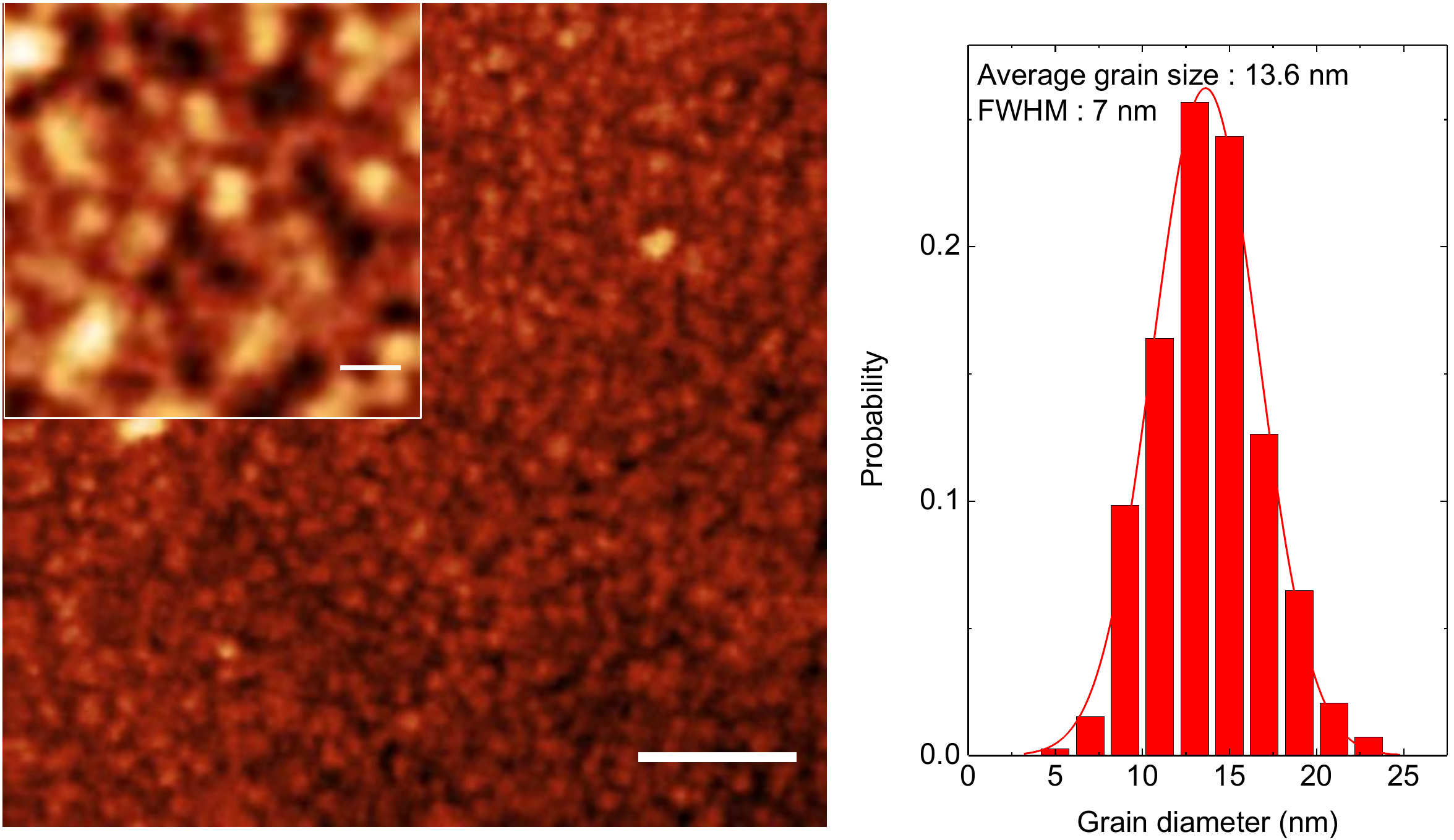}
\caption{AFM image of the sample showing the grain size (scale bar : 100 nm). The inset is a zoom (scale bar : 15 nm). The grain size distribution is shown on the right, with an average grain size of 13.6 nm.}
\label{fig_afm}
\end{figure}

\section{Relation between thickness and domain wall energy distributions}

\begin{figure}[ht]
\includegraphics[width = 0.8\textwidth]{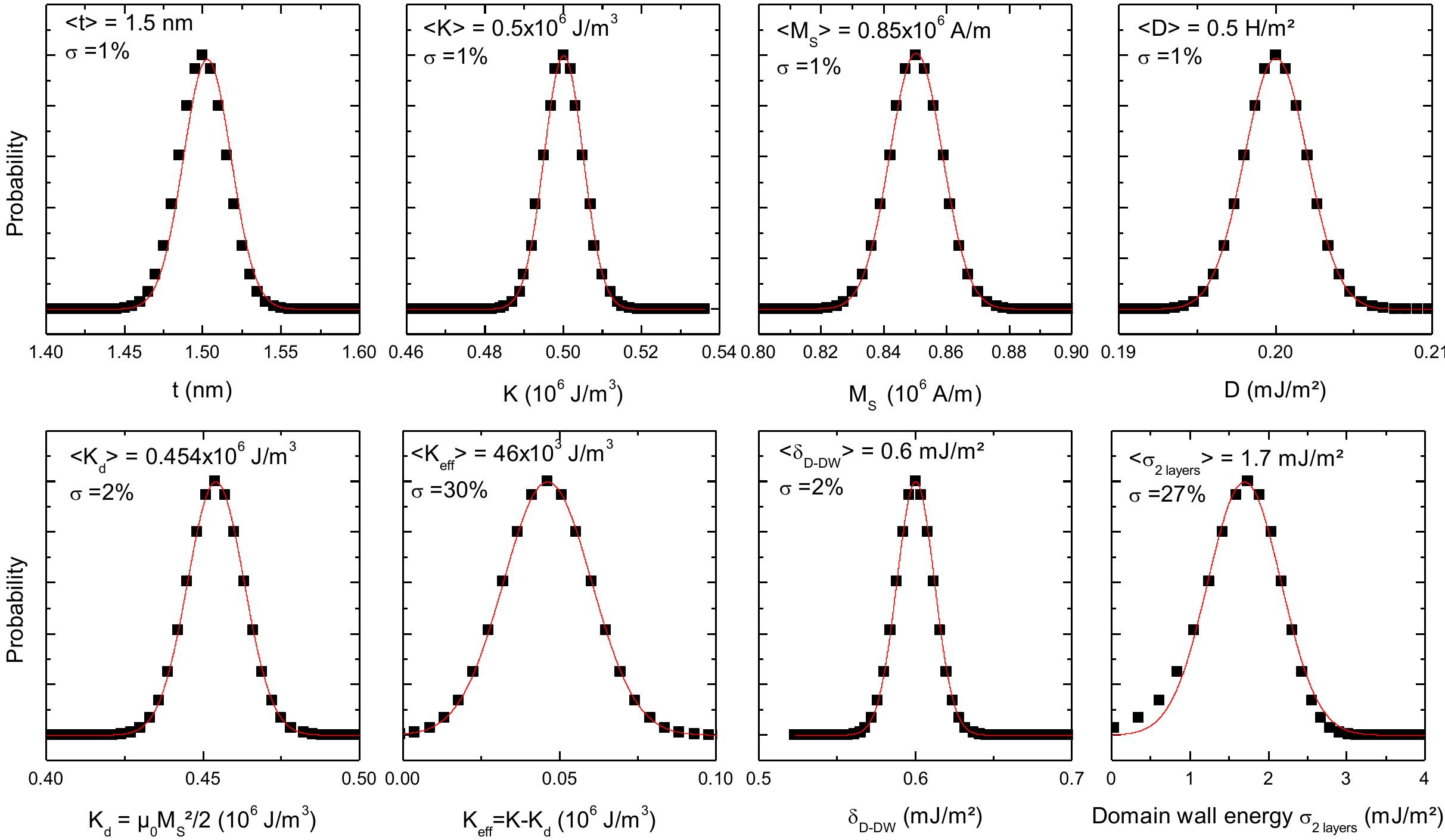}
\caption{Relation between the thickness distribution and the micromagnetic parameters for an initial thickness distribution of 1~\%.}
\label{fig_distri}
\end{figure}

Figure~\ref{fig_distri} shows the distribution of magnetic parameters for an initial thickness distribution with 1\% variance and the averaged material parameters as described in the main text. While the exchange is taken as a constant, the anisotropy $K$ and Dzyaloshinskii-Moriya $D$ constants depend on the thickness due to their interfacial origin as $K = K_S/t$ and $D = D_S/t$ with $K_S$ and $D_S$ the interfacial anisotropy and DMI constants. To account for the grain to grain volume fluctuation in a simulation with constant thickness, the magnetization is varied as $M_S = \langle M_S\rangle t/t_0$. To link these parameters to the skyrmion size, we discuss the domain wall energy distribution \cite{rohart2013}. For a single magnetic film, the domain wall energy is $\sigma_\mathrm{1 layer} \approx 4\sqrt{AK_\mathrm{eff}}-\pi D$ with $K_\mathrm{eff} = K-\frac12\mu_0M_\mathrm{S}^2$ the effective anisotropy, which combines interfacial and shape anisotropy. In our bilayer films, the dipolar couplings add to this formula an other term $\delta_\mathrm{D-DW}$ which further reduces the domain wall energy $\sigma_\mathrm{2 layers}$ \cite{hrabec2016current}. Given its dipolar origin (therefore proportional to $M_\mathrm{S}^2$), it is related to the thickness as $\delta_\mathrm{D-DW} \approx \langle\delta_\mathrm{D-DW}\rangle(t/t_0)^2$, where $\langle\delta_\mathrm{D-DW}\rangle$ has been estimated in Ref.~\onlinecite{hrabec2016current} using micromagnetic simulations to about 0.6 mJ/m$^2$. In a sample suitable for skyrmions, the effective anisotropy is adjusted close to zero, which explains that any variation of the parameters induces a significant variation of the effective anisotropy and therefore the domain wall energy.

\section{Effect of temperature on the size relaxation}

To study how temperature affects the simulated relaxed skyrmion size at a given noise amplitude, we have repeated the simulations with an additional  fluctuating field, which mimics thermal fluctuations at a given temperature\cite{brown1963}. Indeed, in numerical micromagnetism, the temperature is often taken into account by including such a fluctuating field, as described in \cite{Vansteenkiste2014} and references therein. While this approach does reproduce some of the effects of thermal excitations, it must be approached with great care \cite{grinstein2003}. For one, it depends strongly on the cell size, as only the excitations with a wavelength larger than the cell size are included. Furthermore, the material parameters must be adjusted so that the macroscopic (averaged) quantities match the measured properties of the physical system. To limit the extension of these corrections, we have chosen to use a fluctuating field corresponding to a lower temperature than room temperature ($T = 100$ K). The material parameters ($M_\mathrm{S}$, $K$) were adapted so that the spatially-averaged magnetization and the effective anisotropy field $H_\mathrm{K}$ matched the experimental values ($M_\mathrm{S} =869.9$ kA/m instead of 850 kA/m, and $K= 527.66$ kJ/m$^3$ instead of 505 kJ/m$^3$). The exchange stiffness was corrected by twice the relative correction of $M_\mathrm{S}$, as it is an interaction that depends quadratically on the magnetization. In Figure \ref{fig_temperature},
 \begin{figure}[t]
\includegraphics[width = 0.4\textwidth]{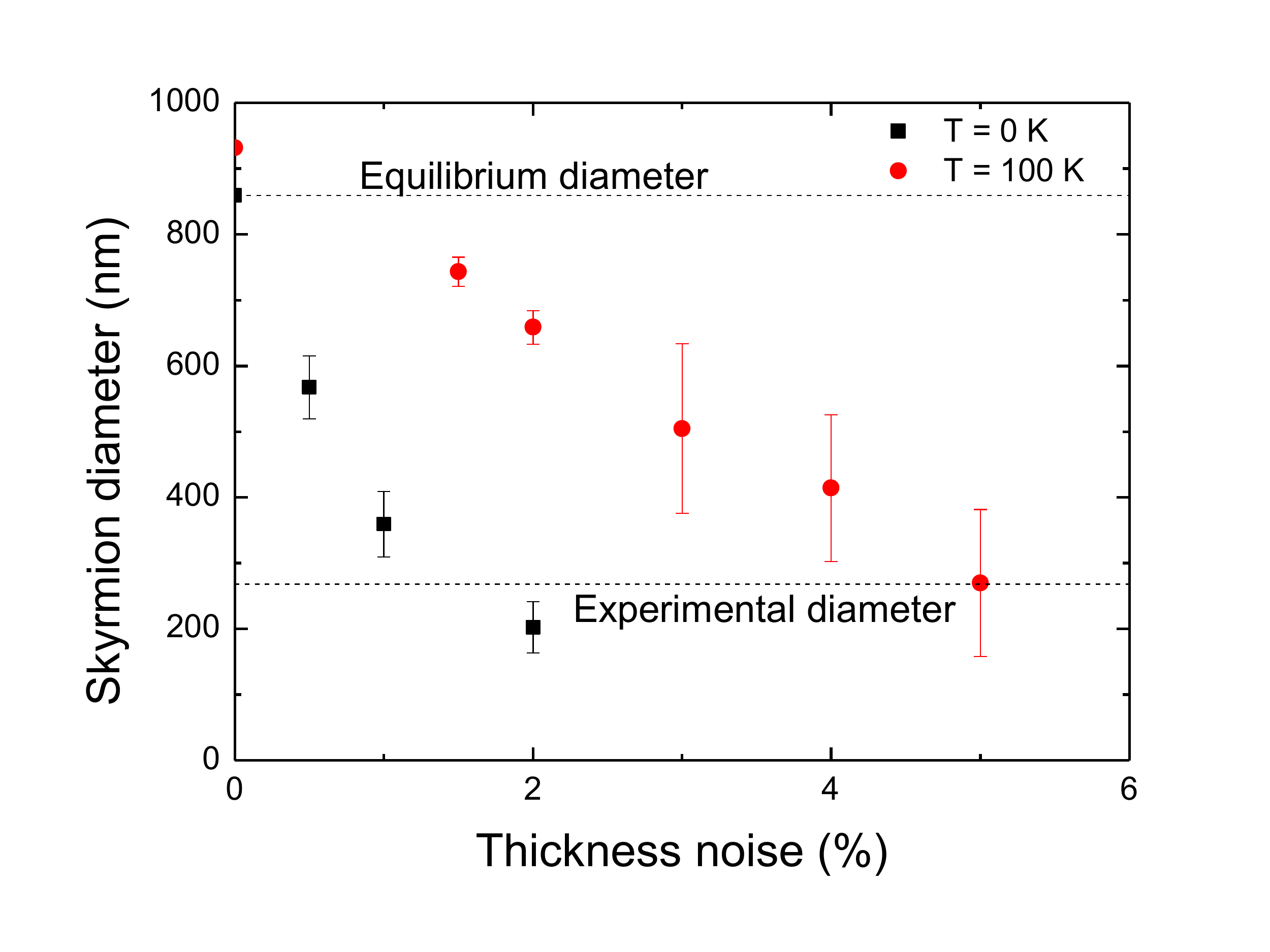}
\caption{Variation of the simulated skyrmion size versus the thickness disorder for 0 (black) and 100 K (red). The dotted lines correspond to the 0~K equilibrium size and the experimental size.}
\label{fig_temperature}
\end{figure}
we show how the relaxed skyrmion diameter varied with the noise amplitude, for $T = 0$ K (black curve; same data as in the main text) and for $T = 100$ K (red curve). Qualitatively, the curves show the same behavior: for lower noise, the diameter approaches the equilibrium value, for larger noise the diameter remains closer to the initial state. For a given noise amplitude, the skyrmions at 100 K are larger, closer to the equilibrium size. Consequently, the noise amplitude that reproduces the experimentally determined diameter (268 nm), is much larger for 100 K than for 0 K (5\% instead of 1-2\%). This result confirms the intuitive expectation that pinning effect is lower for the thermally excited simulation,  as temperature allows overcoming some energy barriers. It indicates that, if indeed the origin of the pinning in the physical system is the variation of grain thickness, the amplitude of this variation is much greater than the 1-2\% we use in the 0 K simulations.

\section{Effect of the grain size}

The impact of the grain size on skyrmion dynamics is essential, as it has been discussed in  previous studies \cite{Reichhardt2015,kim2017current,legrand2017room}. While the grain size used in our study corresponds to the experimental size, we have also considered the effect of different sizes. The same simulation as discussed in the main text (40 nm wide skyrmion relaxed under a 3 mT field, $T= 0$ K) was repeated for different grain sizes, using the same noise amplitude of 1\%. For each grain size, the simulations were repeated with several instances of the random noise and grain shape. The results are shown in Figure\ref{fig_grains}. It can be seen that for smaller grain sizes ($<10$ nm) the obtained skyrmion diameter approaches the equilibrium size, a sign of lower magnetic pinning. For a grain size ranging from 15 nm to 60 nm, the skyrmion diameter seemed to be unaffected by the grain size. This suggest that the relevant length-scale is the DW width (about 13 nm~\cite{hrabec2016current}): grains smaller than the DW width pin less than grains that are bigger or of comparable size. These results are compatible with the findings of Ref.~\onlinecite{kim2017current} for DW pinning.

\begin{figure}[ht]
\includegraphics[width = 0.4\textwidth]{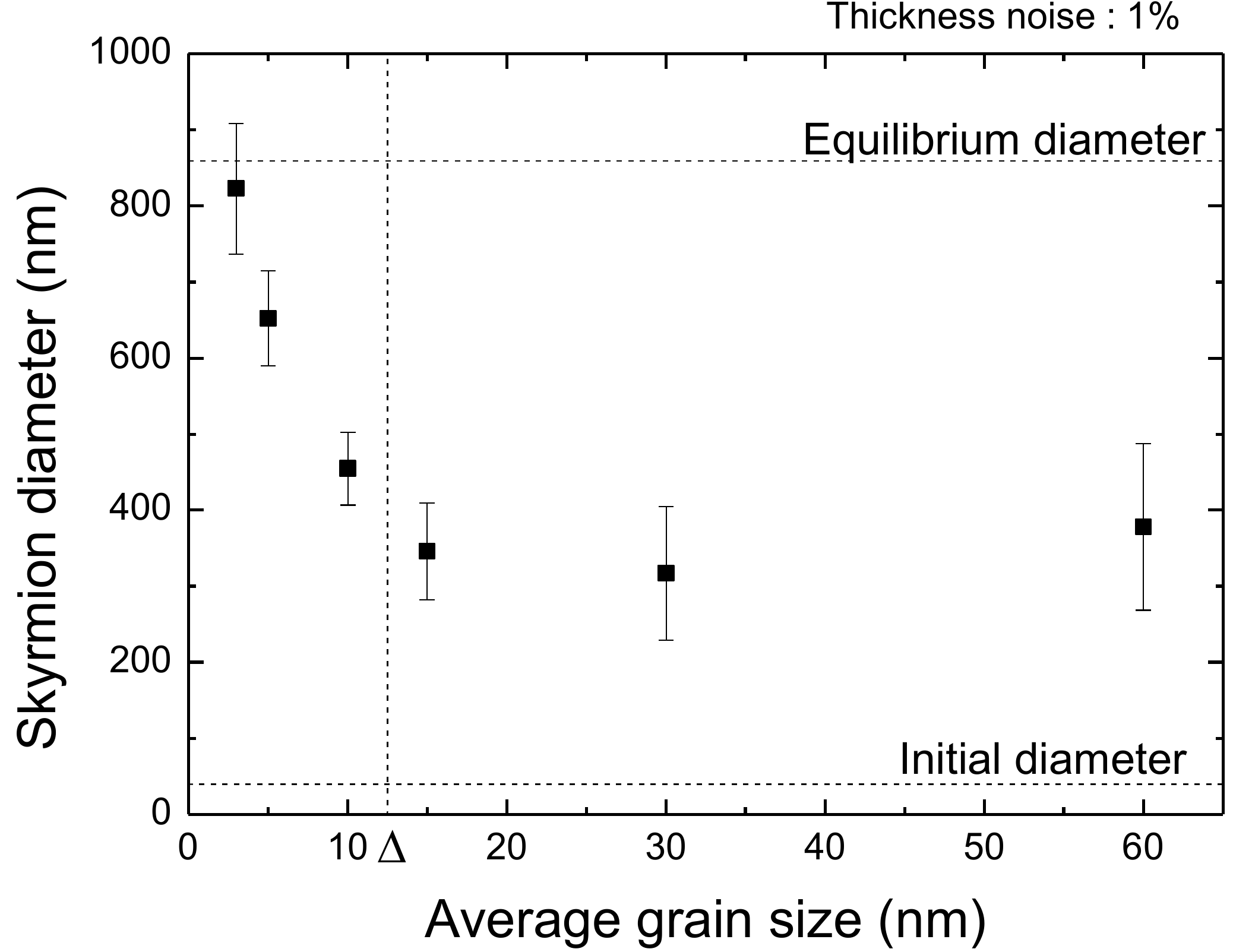}
\caption{Variation of the simulated skyrmion size versus the average grain size. The horizontal lines correspond to the equilibrium, and initial diameters and the vertical line corresponds to the domain wall width parameter $\Delta$.}
\label{fig_grains}
\end{figure}

\end{widetext}

\end{document}